\documentclass[pra,twocolumn,superscriptaddress,noshowpacs]{revtex4-2}
\usepackage{amssymb}
\usepackage{amsmath}
\usepackage{amsfonts}
\usepackage{mathtools}
\usepackage[mathlines]{lineno}
\usepackage{bm}
\usepackage{dcolumn}
\usepackage{physics}
\usepackage{graphicx}
\usepackage[dvipsnames]{xcolor}
\usepackage{color}
\usepackage{xcolor}
\usepackage{calc} % to do for example 0.5\columnwidth+0.2\columnsep
\usepackage{graphics}
\usepackage{epsfig}
\usepackage{epstopdf}
\usepackage{float}
\usepackage{tabularx}
\usepackage{natbib}
\usepackage[colorlinks, linkcolor=blue, urlcolor=blue, anchorcolor=blue, citecolor=blue]{hyperref}
\usepackage{color}
\usepackage{diagbox}
\usepackage{makecell}
\usepackage{multirow}
\usepackage{tabu}
\begin{document}
	
	\title{Circuit QED simulator of two-dimensional Su-Schrieffer-Hegger model: magnetic field induced topological phase transition in high-order topological insulators}
	
	\author{Sheng Li}
	\affiliation{School of Physics, Huazhong University of Science and Technology, Wuhan, 430074, China}
	
	\author{Xiao-Xue Yan}
	\affiliation{School of Physics, Huazhong University of Science and Technology, Wuhan, 430074, China}

	\author{Jin-Hua Gao}
	\email{jinhua@hust.edu.cn}
	\affiliation{School of Physics, Huazhong University of Science and Technology, Wuhan, 430074, China}
	\affiliation{ Wuhan National High Magnetic Field Center,
		Huazhong University of Science and Technology, Wuhan 430074,  China}

	\author{Yong Hu}
	\email{huyong@mail.hust.edu.cn}
	\affiliation{School of Physics, Huazhong University of Science and Technology, Wuhan, 430074, China}

	\begin{abstract}
		High-order topological insulator (HOTI) occupies an important position in topological band theory due to its exotic bulk-edge correspondence. Recently, it has been predicted that external magnetic field can introduce rich physics into two-dimensional (2D) HOTIs. However, up to now the theoretical description is still incomplete and the experimental realization is still lacking. Here we investigate the influence of continuously varying magnetic field on 2D Su-Schriffer-Heeger lattice, which is one of the most celebrated HOTI models, and proposed a corresponding circuit quantum electrodynamics (cQED)  simulator. Our numerical calculation shows that the zero energy corner modes (ZECMs), which can serve as evidence of the high order topology of the lattice, exhibit exotic and rich dependence on the imposed magnetic field and the inhomogeneous hopping strength. Moreover, by exploiting the parametric conversion method, we can establish time- and site-resolved tunable hopping constants in the proposed cQED simulator, thus providing an ideal platform for simulating the magnetic field induced topological phase transitions in 2D HOTIs. Since the high-order topological phases of the proposed model can be characterized by the existence of the ZECMs on the lattice, we further investigate the corner site excitation of the lattice in the steady state limit. Our numerical results imply that the predicted topological phase transitions can be unambiguously identified by the steady-state photon number measurement of the corner sites and their few neighbors. Requiring only current level of technique, our scheme can be readily tested in experiment and may pave an alternative way towards the future investigation of  HOTIs in the presence of magnetic field, disorder, and strong correlation.
\end{abstract}

\maketitle
	\section{Introduction}
\begin{figure*}[tbhp]
		\begin{center}
			\includegraphics[width=1\textwidth]{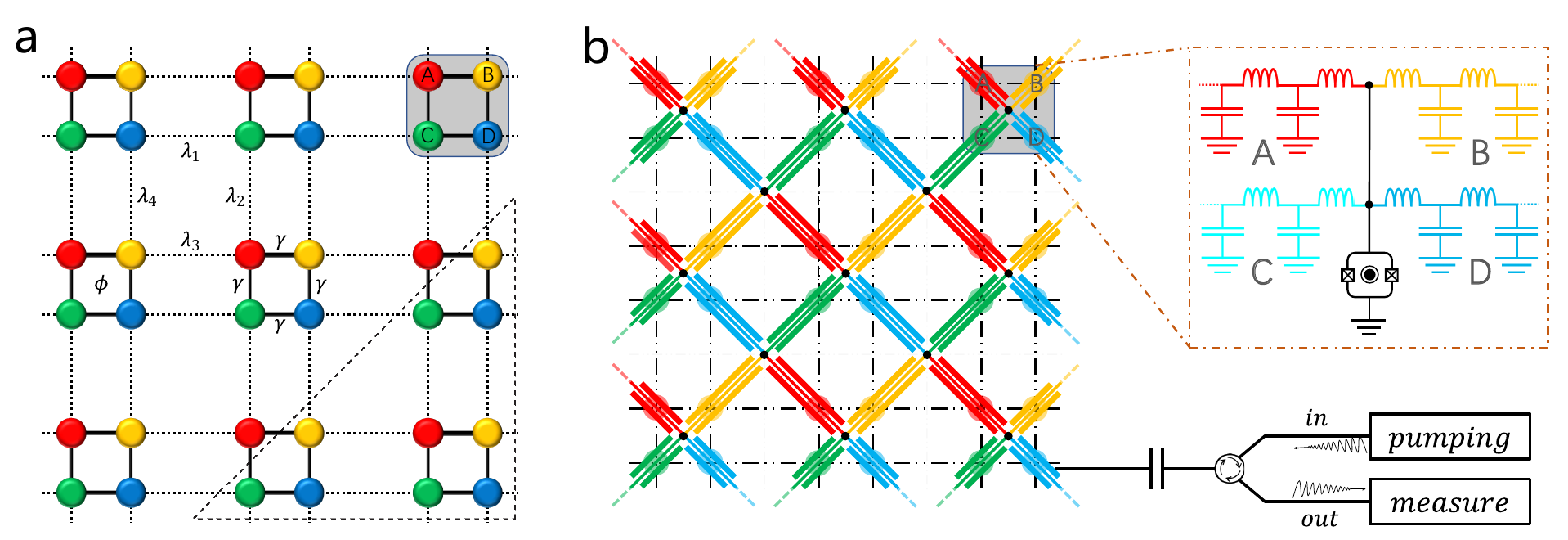}
		\end{center}	
		\caption{\textbf{a} Sketch of the 2D SSH lattice composed of unit-cells with four sites labeled A—D (shadowed area). The lattice sites are coupled through uniform intra-cell coupling strength $\gamma$ and inhomogeneous inter-cell coupling strengths $\lambda_i$. In each intra-cell plaquette, an uniform magnetic flux $\phi$ is penetrated. \textbf{b} Circuit QED simulator of the 2D SSH model. The lattice is built by TLRs grounded at their common ends (big dots) by coupling SQUIDs (crossed squares). The colors of the TLRs label their different eigenfrequencies. In the loop of each SQUID, an externed time-dependent magnetic flux is penetrated to induce the tunable effective coupling between the neighboring TLRs.}
		\label{Fig lattice}
		
	\end{figure*}

	High order topological insulator (HOTI) has attracted extensive research attention in the past few years due to its exotic boundary states \cite{Benalcazar61,Schindlereaat0346}. Generally speaking, a $d$-D $n$th-order HOTI has gapless $(d-n)$-D boundary states, which is noy only determined by but also a representation of its high order band topology \cite{PhysRevLett.119.246402,PhysRevB.96.245115,PhysRevLett.119.246401,PhysRevLett.120.026801,BernevigBook}.
	A typical example is the 2D Su-Schrieffer-Heeger (SSH) lattice shown in Fig.~\ref{Fig lattice}a, which has four-site unit-cells coupled through homogeneous intra-cell hopping strength $\gamma$ and inhomogeneous inter-cell hopping strengths $\lambda_i$ \cite{PhysRevB.98.201114,Benalcazar61,PhysRevLett.118.076803,kim2019topological}. This lattice can host 0D zero-energy corner modes (ZECMs) instead of 1D gapless edge modes under certain circumstances, thus belonging to 2D second-order HOTI. Recently, the influence of external magnetic field on 2D HOTI has been investigated, and a variety of novel topological phases have been predicted \cite{PhysRevB.100.245108,Benalcazar61}. The motivation and the physics behind is that the imposed magnetic field can change the symmetry of the Hamiltonian \cite{PhysRevLett.125.236804}. It has been pointed out that the high-order topology of the 2D SSH model depends on both the uniformly-imposed plaquette magnetic field $\phi$ and the ratio $\lambda_i/\gamma$ (Fig.~\ref{Fig lattice}a). For $\lambda_i \textgreater \gamma$ and  $\phi=\pi$ \cite{Benalcazar61,PhysRevB.100.085138}, this model enters the topological non-trivial region characterized by the appearance of ZECMs \cite{PhysRevResearch.3.013239}. However, theoretical research up to now has focused only on several discrete values of the imposed magnetic field, e.g., $0$-flux and $\pi$-flux in Ref.~\cite{Benalcazar61} and $2\pi/3$-flux and $2\pi/10$-flux in Ref.~\cite{PhysRevB.100.245108}. The detailed and complete description of the role of magnetic field in HOTI is still lacking. From this point of view, building a corresponding analog quantum simulator \cite{RevModPhys.86.153} and investigating the behavior of the boundary states \cite{TopoPhoton2019RMP,LuTPReviewNatPhoton2014} can offer insight and inspiration for further research. Meanwhile, despite the experimental progress of realizing HOTIs in various metamaterials \cite{PhysRevB.99.201411,PhysRevLett.122.233903,PhysRevLett.122.233902,PhysRevB.98.205147,2018Observation,RN368,RN370,PhysRevB.99.020304}, the proposed magnetic-field-induced topological phase transition has not been experimentally demonstrated yet, partially because the simultaneous realization of the site-dependent inhomogeneous hopping strength and the tunable synthetic gauge field is still challenging.
	
	On the other hand, superconducting quantum circuit (SQC) has been regarded as a promising platform of realizing quantum simulation \cite{RN458,2009}. The lattice sites in this setup are constructed by superconducting transimissionline resonators (TLRs) \cite{RevModPhys.93.025005} and superconducting qubits, and electrons in conventional materials are replaced by microwave photons. Compared with other physical systems, SQC takes the advantage of flexible engineering of a large scaled lattice, which stems from the tunability and scalability in the circuit design and control \cite{RN443}. With advances in theory and technology, the controllable coupling between SQC elements has been experimentally demonstrated \cite{PhysRevA.90.022307,PhysRevLett.127.080505}, leading to the on-demand synthesization of artificial gauge field \cite{RN377} and many-body localization \cite{Roushan1175}. Moreover, the strong coupling between SQC elements \cite{RevModPhys.93.025005,LiuYXReview2017PR} allows the effective Kerr type \cite{HartmannCoupledCavity2006NP} and Jaynes-Cummings-Hulbard type \cite{GreenTreeJCHNP2006} nonlinearity and consequently the further simulation of strongly correlated photonic liquids \cite{PhysRevLett.103.033601,PhysRevLett.108.223602,2009,RevModPhys.85.299,RN444}.

	In this manuscript, we study the high order topology of 2D SSH lattice induced by continuously varying magnetic field and its quantum simulation in SQC system. In the first step, we propose a flexible circuit quantum electrodynamics (cQED) quantum simulator of 2D SSH lattice. Here the lattice sites are constructed by superconducting TLRs and coupled to their neighbors by grounding superconducting quantum interference devices (SQUIDs) (Fig.~\ref{Fig lattice}b). The photon hopping on this lattice is induced through the parametric modulation of the grounding SQUIDs \cite{RN373,RN377,WangYPNPJQI2016,PhysRevA.93.062319}. The distinct merit of this scheme is that it can provide time- and site-resolved tunability of both the hopping strengths and the hopping phases between the adjacent sites, thus enabling the simultaneous establishment of tunable inhomogeneous hopping strengths which is critical for the lattice to have non-trivial high order topology, and the non-trivial hopping phases which lead to controllable synthetic magnetic field for microwave photons.
	
	The tunability of the proposed simulator in turn stimulates us to go beyond previous work \cite{PhysRevB.100.245108,Benalcazar61} and calculate the energy spectrum of the 2D SSH lattice versus continuous external magnetic field and the ratio between the intra- and inter-cell hopping strengths. Our results show that the spectrum and the ZECMs (and thus the high order topology of the lattice) exhibit rich and complicated behavior. In particular, the HOTI phase diagram of the lattice can be generally divided into five parts (see detailed discussion in Sec.~\ref*{S}) based on the existence of the ZECMs. In certain phases regions, the non-trivial HOTI phase is robust against the imposed magnetic field, while in other regions the imposed magnetic field can lead to band gap closing and thus induce HOTI phase transition. Our estimation based on recently reported experimental data \cite{RN373,NISTHongOuMandelPRL2012,RN377,PhysRevLett.127.100503,DCEexperimentNature2011} further pinpoints that the parameters of all the predicted HOTI phase regions can in principle be achieved in the proposed cQED simulator.
	
	 %With the proposed simulator, our numerical results imply that the HOTI phase exhibit complicated depending on the magnetic field $\phi$ and the ratio $\lambda_i/\gamma$. As $\gamma$ passes the cirtical point, pinpoint a HOTI phase transition occuring at $\phi=\pi/2$.
	
	 As the high order band topology of the proposed model can be characterized by the ZECMs, we further investigate the discrimination of the predicted HOTI phases through the corner-site pumping and the consequent steady-state photon number (SSPN) distribution measurement. The essential physics is that the spatial and spectral localization of the ZECMs can lead to exotic spatially localized SSPN distribution, and the latter can be used as strong evidence of the existence of the ZECMs. Our numerical results imply that with current level of technology, one can clearly identify the existence of the ZECMs and consequently the boundaries of the predicted HOTI phase regions by only measuring  few sites near the corner.
	%Circuit parameters estimated from reported experimental data indicates that all the predicted magnetic-field-induced HOTI phases of the 2D SSH lattice can in principle be achieved. As the band topology of the proposed model can be recognized by the occurrence and absence of the ZECMs, we further investigate the discrimination of the predicted topological phases by numerically calculating the corner-site pumping and the consequent steady-state photon number distribution(SSPN).
	%After proposing the cQED lattice with model parameters estimated based on reported experimental data, we then  discuss how to detect the corner modes of the lattice via  the pumping and the steady-state photon number (SSPN) measurements, because the emergence of the corner modes pinpoint the occurrence of HOTI. %
	%Two numerical simulations are explicitly given. The first case is  $\lambda_i \textgreater \gamma$ with magnetic $\phi \in \left[0, \pi\right]$. Our results indicate the disappearance of the ZECMs at $\phi \leq \pi/2$, which is in accordance with the predicted topological phase transition at $\phi=\pi/2$ \cite{PhysRevLett.125.236804}, and can be clearly identified by the SSPN measurement around the corner sites.
	%which is in good agreement with the theoretical analysis from the butterfly spectrum.%
	%The other situation is fixed $\phi=\pi$ flux with varying anisotropic $\lambda_i$, in which two distinct HOTI phases characterzied by two-ZECMs and four-ZECMs can be unambiguously observed. From this point of view,
	Our cQED architecture can therefore be regarded as an efficient platform of investigating the magnetic field effect on 2D HOTIs. Due to the flexibility and scalability offered by the SQC system, our proposal can be readily generalized to other HOTI lattice configurations \cite{PhysRevLett.122.086804,PhysRevLett.120.026801}, and can allow the further incorporation of non-Hermicity \cite{Ashida,doi:10.1080/00018732.2021.1876991}, disorder \cite{UnderwoodDisorderPRA2012}, and strong correlation \cite{PhysRevLett.103.033601,PhysRevLett.108.223602,2009,RevModPhys.85.299,RN444}, thus providing an alternative route of exploring HOTI in the future.
	
	\section{Circuit QED lattice simulator of 2D SSH model}
	
	As shown in Fig.~\ref{Fig lattice}b, the proposed cQED lattice simulator is built up by four kinds of TLRs differed by their eigenfrequencies and placed in an interlaced square pattern \cite{WangYPNPJQI2016}. These four kinds of TLRs  correspond to the A-D sublattice sites in Fig.~\ref{Fig lattice}a, respectively. In addition, we ground these TLRs at their common ends by grounding  SQUIDs with effective Josephson inductances much smaller than those of the TLRs \cite{RN373,NISTHongOuMandelPRL2012}. The roles of the grounding SQUIDs are two-fold. Firstly, their small inductances impose shortcut boundary conditions for the TLRs. Due to the current dividing mechanism, a current flowing away from a particular TLR will prefer flowing to ground rather than to its neighbors \cite{PhysRevA.90.022307}. Therefore, we can exploit  the lowest $\lambda/2$ eigenmodes of the TLRs as the uncoupled localized Wannier modes of the lattice, and write the on-site part of the lattice Hamiltonian as
	\begin{equation}
	\mathcal{H}_{\mathrm{S}}=\sum_{\alpha,\mathbf{r}} \omega_{\alpha} \alpha_{\mathbf{r}}^{\dagger} \alpha_{\mathbf{r}},
	\end{equation}
	where ${\alpha^{\dagger}}_{\mathbf{r}}/\alpha_{\mathbf{r}}$ are the creation/annilhilation operators of the $\alpha$th site in the $\mathbf{r}$th unit-cell for $\alpha \in { A,B,C,D}$, and $\omega_{\alpha}$ are the corresponding eigenfrequencies. For the following establishment of the parametric frequency conversion (PFC) process, we further specified $\omega_{\alpha}$ as $(\omega_{A},\omega_{B},\omega_{C},\omega_{D})=(\omega_{0},\omega_{0}+\Delta,\omega_{0}+4\Delta,\omega_{0}+3\Delta)$ with $\omega_{0}/2\pi\in[6,10]\mathrm{GHz}$ and $\Delta/2\pi\in[0.5,1]\mathrm{GHz}$. With current level of technology, such configuration can be realized with very high precision through the design and fabrication of the circuit (e.g. length selection or impedance engineering) \cite{UnderwoodDisorderPRA2012,PhysRevLett.127.100503}.
	
	The second function of the grounding SQUIDs is to  implement the effective Hamiltonian
	\begin{equation}
	\mathcal{H}_{\mathrm{T}}=\sum_{\left\langle(\mathbf{r}, \alpha),\left(\mathbf{r}^{\prime}, \beta\right)\right\rangle} \mathcal{T}_{\mathbf{r},\alpha}^{\mathbf{r}^{\prime},\beta} \beta_{\mathbf{r}^{\prime}}^{\dagger} \alpha_{\mathbf{r}} e^{i \theta_{\mathbf{r}, \alpha}^{\mathbf{r}^{\prime}, \beta}}+\mathrm{H.C.},
	\label{Eq HoppingHamiltonian}
	\end{equation}
	in the rotating frame of $\mathcal{H}_\mathrm{S}$ through the PFC method. Here $\mathcal{T}_{\mathbf{r},\alpha}^{\mathbf{r}^{\prime},\beta}$ labels the real $(\mathbf{r}, \alpha) \Leftrightarrow \left(\mathbf{r}^{\prime}, \beta\right)$ hopping strength sketched in Fig.~\ref{Fig lattice}a, and $\theta_{\mathbf{r}, \alpha}^{\mathbf{r}^{\prime}, \beta} = \int_{\mathbf{r}, \alpha}^{\mathbf{r}^{\prime}, \beta}\mathrm{d}\textbf{x} \cdot\textbf{A}(\textbf{x})$ is the corresponding hopping phase manifesting the existence of the vector potential $\textbf{A}(\textbf{x})$ \cite{GoldmanGauge2014RPP}. We establish Eq. \eqref{Eq HoppingHamiltonian} through the dynamic modulation of the grounding SQUIDs \cite{WangYPNPJQI2016,RN377,PhysRevLett.127.100503}. The physics can be briefly illustrated in the following two steps:
	\begin{enumerate}
		\item Let us consider a particular neighboring TLR pair $\left\langle(\mathbf{r}, \alpha),\left(\mathbf{r}^{\prime}, \beta\right)\right\rangle$. Due to the very small inductances of their common grounding SQUID, the voltage across the SQUID is very small, and the grounding SQUID works effectively as a linear inductance which can be tuned by its penetrated flux.  As the currents of the two TLRs flow through the same grounding SQUID, an inductive current-current coupling
		\begin{equation}
		\mathcal{H}_\mathrm{S}^{\mathbf{r}\alpha,\mathbf{r}^{\prime}\beta}= \mathcal{T}^{\mathrm{ac}}_{\mathbf{r}\alpha, \mathbf{r^{\prime}}\beta}(t)(\alpha_{\mathbf{r}}+\alpha_{\mathbf{r}}^{\dagger}) (\beta_{\mathbf{r^{\prime}}}+\beta_{\mathbf{r^{\prime}}}^{\dagger}),
		\end{equation}
		can be induced, with the time-dependent coupling constant $\mathcal{T}^{\mathrm{ac}}_{\mathbf{r}\alpha, \mathbf{r^{\prime}}\beta}(t)$ proportional to the Josephson coupling energy (and consequently controlled by the penetrated flux) of the grounding SQUID. We then a.c. modulate the grounding SQUID with frequency bridging the frequency gap between the two TLRs. With this modulation, a PFC process between the TLRs can be induced, described in the rotating frame of $\mathcal{H}_\mathrm{S}$ by
		\begin{equation}
		\mathcal{H}_{\mathrm{T}}^{\mathbf{r} \alpha,\mathbf{r}^{\prime}\beta}=\mathcal{T}_{\mathbf{r},\alpha}^{\mathbf{r}^{\prime},\beta} \beta_{\mathbf{r}^{\prime}}^{\dagger} \alpha_{\mathbf{r}} e^{i \theta_{\mathbf{r}, \alpha}^{\mathbf{r}^{\prime}, \beta}}+\mathrm{H.C.},
		\end{equation}
		where the hopping amplitude $\mathcal{T}_{\mathbf{r},\alpha}^{\mathbf{r}^{\prime},\beta}$ is proportional to the amplitude of the modulating tone, and the hopping phase $\theta_{\mathbf{r}, \alpha}^{\mathbf{r}^{\prime}, \beta}$  is exactly the initial phase of the modulating tone. If we want to induce the on-site term, we merely need to adjust the modulating frequency and choose  a corresponding different rotating frame. With reported experimental data \cite{RN373,NISTHongOuMandelPRL2012,RN377,PhysRevLett.127.100503,DCEexperimentNature2011}, the hopping strength $\mathcal{T}_{\mathbf{r},\alpha}^{\mathbf{r}^{\prime},\beta}$ can further be estimated as  $\mathcal{T}_{\mathbf{r},\alpha}^{\mathbf{r}^{\prime},\beta} \in \left[ 5, 15 \right] \mathrm{ MHz}$.  Moreover, previous discussions \cite{WangYPNPJQI2016,PhysRevA.93.062319} have suggested that such PFC formalism is robust against the fabrication error and the $1/f$ noise in SQC \cite{FlickerRMP2014}, implying that the $\mathcal{H}_{\mathrm{T}}^{\mathbf{r} \alpha,\mathbf{r}^{\prime}\beta}$ term can be stably and precisely synthesized.
		
		\item We then generalize the described PFC method to each of the hopping link on the lattice: We modulate each of the grounding SQUIDs on the lattice with pulses containing three tones $\Delta$, $2\Delta$, and $4\Delta$. A close inspection indicates that we can independently control every vertical hopping branch and every pair of horizontal hopping branches by a modulating tone threaded in one of the grounding SQUIDs \cite{WangYPNPJQI2016}. The summation of all these hopping terms thus leads to the establishment of the effective Hamiltonian in Eq. \eqref{Eq HoppingHamiltonian} with time- and site-resolved tunability. In particular, the site-dependent hopping strength meets the requirement of inhomogeneous hopping amplitudes, which is important for the 2D SSH lattice to enter the topological non-trivial region \cite{PhysRevLett.119.246402,PhysRevB.98.205422}, while the controllable hopping phases pinpoint  the creation of arbitrary synthetic magnetic field with Landau gauge for microwave photons.
	\end{enumerate}

	%	\section{Measurement of the corner states}

	\begin{figure*}[tbhp]
		\begin{center}
			\includegraphics[width=1\textwidth]{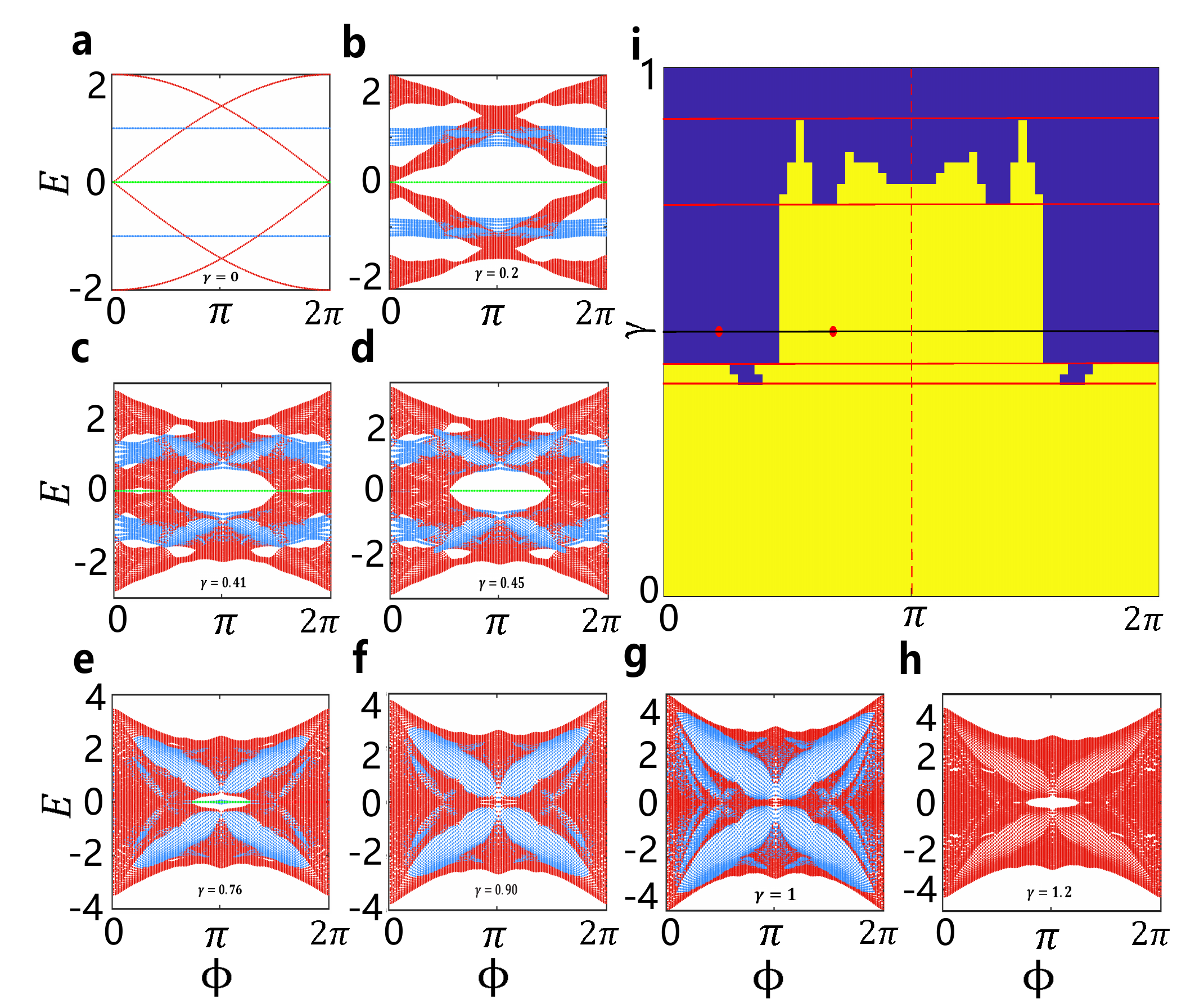}
		\end{center} 	
		\caption{\textbf{a} - \textbf{h } Numerically simulated HBS under OBC for varying $\gamma$. The bulk, edge, and corner modes are labelled by the red, bule, and green lines, respectively. \textbf{i} HOTI phase diagram of the lattice. Here we characterize the topological trivial/non-trivial region simply by the absence/existence of the ZECMs, and label them by the blue/yellow colors, respectively. The phase diagram is mirror-symmetric with respect to $\phi=\pi$ (the red dashed line). The red solid lines denote the gap closing points $\gamma_{c1}=0.41$, $\gamma_{c2}=0.45$, $\gamma_{c3}=0.76$, and $\gamma_{c4}=0.90$ from bottom to top (see main text for details). The black line with red dots at $\phi=2\pi/3$ and $\phi=2\pi/10$ is $\gamma=0.5$ which is used for the numerical simulation in Sec.~\ref*{fig}.}
		\label{Fig butterfly}
	\end{figure*}

	%	\section{Circuit QED simulator of 2D SSH with tunable artificial magnetic field and hopping strength}
	
	\section{High order topological phases versus varying magnetic field in 2D SSH lattice}
	\label{S}
	Currently the detailed-and-ultimate characterization of high order band topology in the presence of magnetic field is still under debate \cite{PhysRevB.100.245108,Benalcazar61,PhysRevLett.125.236804,PhysRevB.100.085138,PhysRevLett.122.086804,PhysRevLett.120.026801}. Motivated by the principle of bulk-edge correspondence \cite{BernevigBook}, we follow an alternative route of investigation in this manuscript, that is,  to study the boundary behavior of the lattice.
	In the rest of this manuscript, we consider the corner excitation physics on a cQED lattice consisting of $8 \times 8$ unit-cells (i.e. $16 \times 16$ sites).
	In the first situation, we consider the uniform inter-cell hopping and set $\lambda_i$ as unity, i. e. $\lambda_i=1$ for $i=1,2,3,4$.
	The intra-cell hopping strength $\gamma$ and the magnetic flux $\phi$ is set in the range $\gamma \in \left[0,1.2 \right]$ and $\phi \in \left[0,2\pi \right]$, respectively.
	The Hofstadter butterfly energy spectrum (HBS) under open boundary condition (OBC) is calculated and shown in Fig.~\ref{Fig butterfly} (the truncation is placed between the unit-cells).
	Compared with that under periodic boundary condition, the HBS under OBC has edge modes (blue lines) and ZECMs (green lines) emerged from the bulk modes (red lines).
	In what follows,  we characterize the topology of the considered lattice by the behavior of the ZECMs.
	As shown in Fig.~\ref{Fig butterfly}, the HBS and the topological phase of the lattice versus $\gamma$ and $\phi$ can be generally divided into five situations (here we only describe the range $\phi \in \left[0,\pi \right]$ because the HBS exhibits mirror symmetry with respect to $\phi=\pi$):
	
	\begin{enumerate}
		\item As shown in Fig.~\ref{Fig butterfly}a, The HBS at $\gamma=0$ can be classified into bulk, edge, and corner state. The eigenvalues of the edge states are 28-fold degenerate in our $8\times 8$ unit-cells, with the degeneracy depending on the lattice size, while the zero eigenvalue of the corner states are four-fold degenerate, with the degeneracy irrelevant of the lattice size. The existence of the ZECMs pinpoints that the 2D SSH model now is a non-trivial HOTI. As $\gamma$ increases in the range $\gamma\in (0,0.41)$ (Fig.~\ref{Fig butterfly}b), the degeneracy of the edge states is destroyed and the energy band of the edge states becomes broad.
		Meanwhile,  the corner states still remain zero-energy and four-fold degenerate. The band gap at $E=0$ between the bulk modes and the ZECMs exists in all range of $\phi$ and decreases with increasing $\gamma$, implying that the HOTI phase now is robust against the imposed magnetic fields.
		Here the upper bound $\gamma_{c1}=0.41$ (and the lower and upper bounds in the other situations) is not an analytic result but obtained numerically.
		\item As $\gamma$ passes the point $\gamma_{c1}=0.41$ and varies in the range $\gamma \in \left[ 0.41,0.45 \right]$, the band gap and consequently the ZECMs always exist for $\phi \in [\pi/2,\pi]$.
		However, the band gap is closed at points located in $\phi \in [0,\pi/2]$ (Fig.~\ref{Fig butterfly}c). Here the gap-closing is not complete in the sense that the gap is not closed for all $\phi \in [0,\pi/2]$.
		The ZECMs in this range behaves much more complicated: the ZECMs still exist for those $\phi$ where the band gap is still open.
		Meanwhile, for those $\phi$ where the band gap is closed, the ZECMs may either disappear or even co-exist with the bulk modes.
		\item After $\gamma$ goes across the point $\gamma_{c2}=0.45$  and moves into the range $\gamma \in [0.45,0.76]$, the band gap closes completely in the range $\phi \in \left[0, \pi/2 \right]$ but still exist in the whole range of $\phi \in \left[\pi/2,\pi \right]$, as shown in Fig.~\ref{Fig butterfly}d.
		The band width of the edge state now becomes larger, and approaches the zero-energy value in the range $\phi \in \left[0, \pi/2 \right]$.
		Meanwhile, the ZECMs exist in the whole range of $\phi \in \left[\pi/2,\pi \right]$ and disappear completely in the whole range of $\phi \in \left[0,\pi/2 \right]$.
		It implies that the system with $0\textless\phi\textless\pi/2$ can not be continuously deformed into that with $\pi/2\textless\phi\textless\pi$ due to the gap-closing at the $\phi=\pi/2$ \cite{PhysRevLett.125.236804,PhysRevB.100.245108}, and pinpoint a HOTI phase transition occuring at $\phi=\pi/2$.
		\item For $\gamma$ in the range $\gamma \in [0.76,0.90]$, the band gap closes completely in the range $\phi \in [0,\pi/2]$ and incompletely in $\phi \in [\pi/2,\pi]$.
		With increasing $\gamma$, the band gap shrinks continuously and eventually vanishes at $\gamma_{c4}= 0.90$ (Figs.~\ref{Fig butterfly}e and~\ref{Fig butterfly}f).
		\item Finally, in the range $\gamma >0.90$, the band gap near $E=0$ is reopened in the range $\phi \in [\pi/2,\pi]$. The ZECMs, however, disappear. The absence of the ZECMs imply that now the lattice is in the topological trivial phase (Figs.~\ref{Fig butterfly}g and~\ref{Fig butterfly}h).
	\end{enumerate}
	
	As a summary of the above results, we futher plot in Fig.~\ref{Fig butterfly}i the HOTI phase diagram of the lattice. Notice that our result is consistent with previous work which considered only several discrete values of $\phi$ \cite{PhysRevB.100.245108}. Here we label the trivial/non-trivial HOTI phases simply by the existence of the ZECMs.

	\begin{figure*}[tbhp]
		\begin{center}
			\includegraphics[width=0.9\textwidth]{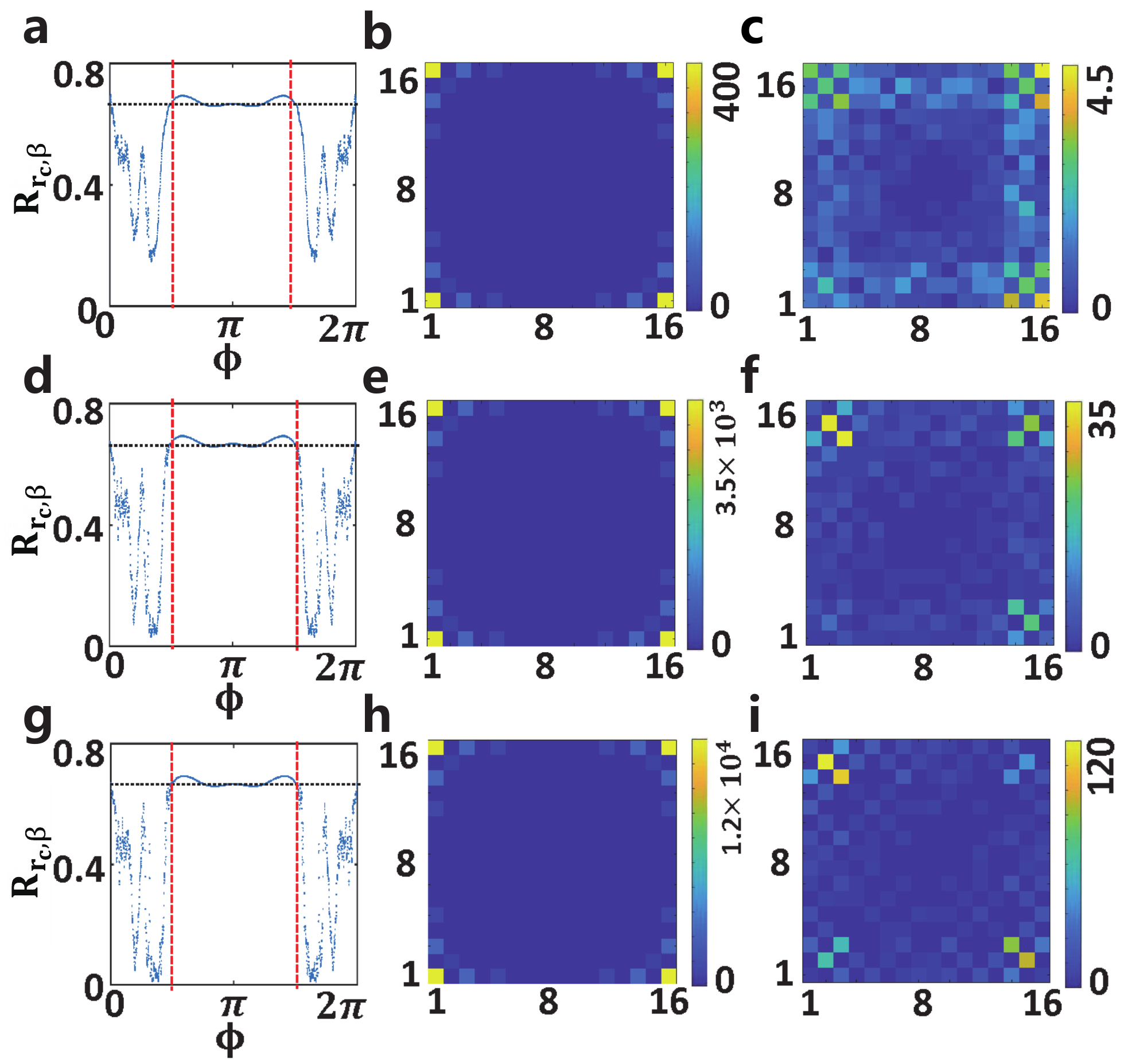}
		\end{center} 	
		\caption{Numerically simulated SSPN distributions versus magnetic field $\phi$ and dissipation factor $\kappa$. \textbf{a} depicts the concentration factor $R_{\mathbf{r}_\mathrm{c},\beta}$ at $\kappa=0.03$. Here the summation $N_{\mathbf{r}_\mathrm{c},\beta}$ is performed at the corner sites and their nearest six neighbors indicated by the dashed triangle in Fig.~\ref{Fig lattice}a. The red vertical dashed lines denote the topological phase transition points $\phi=\pi/2$ and $\phi=3\pi/2$, and the black horizontal dotted line labels the empirical critical value $R_{\mathbf{r}_\mathrm{c},\beta} = 0.7$. The SSPNs distribution of the whole lattice at $\phi = 2\pi/3$ and $\phi = 2\pi/10$ are sketched in \textbf{b} and \textbf{c}, respectively. \textbf{d}-\textbf{f} and \textbf{g}-\textbf{i} show the same calculations at $\kappa=0.01$ and $\kappa=0.005$, respectively.}
		\label{Fig measurementfield}
	\end{figure*}
	\section{Measurement scheme of the proposed HOTI phase}
	\label{fig}
	In the next step, we consider the measurement scheme of the proposed topological phases. With current level of technology, the measurement of HBS of small-size lattice has been reported \cite{Roushan1175}. However, this method can hardly be generalized to large-scale lattice, because the energy spacing becomes much more dense with increasing lattice size. Therefore, we consider the observation scheme based on the corner site excitation of the proposed lattice. The motivate is that the proposed high order topology is characterized by the behavior of the ZECMs and the ZECMs are spatially localized near the corners and spectrally located at $E = 0 $. We can capacitively couple external pump coil to the corner sites (Fig.~\ref{Fig lattice}b) and inject pulses matching the zero frequency in the rotating frame of $\mathcal{H}_\mathrm{S}$. If the ZECMs exist, the response of the lattice will accordingly exhibit spatially localized patterns, otherwise the response will extend over a wide range on the lattice. In this sense, we can determine the existence of the ZECMs and consequently the topology of the lattice by measuring merely few sites near the corners. Two candidate physical observables can be considered. The first one is the SSPN: We pump the corner sites for sufficiently long time and let the lattice approach its steady state. Then the SSPN of the corner sites and its few neighbors can be determined by measuring the energy leaking out (Fig.~\ref{Fig lattice}b). The second one is the reflection coefficient: We input a pulse into the corner site and measure the output photon currents from the corner sites and their neighbors. Based on the input-output formalism \cite{RevModPhys.82.1155}, the spatial and spectral distribution of the ZECMs can be determined from the reflected signals. Both these two schemes have been reported in recent experiments \cite{doi:10.1063/1.4919759,PhysRevX.7.041043}. In the following, we perform numerical simulations following the SSPN scheme: The coherent monochromatic pumping of a particular single-site $(\mathbf{r}_{\mathrm{p}},\alpha)$ can be described by
	\begin{equation}
	\mathcal{H}_{\mathrm{pump}} = \mathcal{P}^{\dagger} \mathbf{a} e^{-i \Omega_{\mathrm{P}} t}+\mathrm{h} . \mathrm{c}. ,
	\end{equation}
	where $\mathcal{P}$ is the pumping strength vector with only one non-zero element at $(\mathbf{r}_{\mathrm{p}},\alpha)$, $\mathbf{a}$ is the vector of annilhilation operators of the lattice sites, and $\Omega_{\mathrm{P}}$ is the detuning of the pumping frequency in the rotating frame of $\mathcal{H}_\mathrm{S}$. The SSPN of the lattice can be obtained through the equation \cite{PhysRevA.93.062319}
	\begin{equation}
	i \frac{\mathrm{d}\langle\mathbf{a}\rangle}{\mathrm{d} t}=\left[\mathcal{B}-\left(\Omega_{\mathrm{P}}+\frac{1}{2} i \kappa\right) \mathcal{I}\right]\langle\mathbf{a}\rangle+\mathcal{P}=0,
	\label{1}
	\end{equation}
	where the matrix $\mathcal{B}$ is defined by $\mathbf{a}^\dagger\mathcal{B}\mathbf{a}=\mathcal{H}_{\mathrm{T}}$. Without loss of generality here we assume that the TLRs have uniform decay rate $\kappa$.
	
	Suppose $(\mathbf{r}_{\mathrm{c}},\beta)$ is a corner site, we can measure the SSPN $n_{\mathbf{r}_\mathrm{c},\beta} = \left\langle\beta_{r_{c}}^{\dagger} \beta_{r_{c}}\right\rangle$ at exactly this site, and the summation of SSPN at $(\mathbf{r}_{\mathrm{c}},\beta)$ and its few neighbors, which we denote as $N_{\mathbf{r}_\mathrm{c},\beta}$. Then the concentration factor $R_{\mathbf{r}_\mathrm{c},\beta}= n_{\mathbf{r}_\mathrm{c},\beta}/N_{\mathbf{r}_\mathrm{c},\beta}$ serves as a good index of the existence of the ZECMs. A significant high $R_{\mathbf{r}_\mathrm{c},\beta}$ can be obtained only if the following three requirements are met simultaneously: 1. the ZECMs do exist, i.e. the lattice is in its topological non-trivial phase. 2. $(\mathbf{r}_{\mathrm{c}},\beta)=(\mathbf{r}_{\mathrm{p}},\alpha)$, i.e. what we pump is a corner site. 3. $\Omega_{\mathrm{P}}\approx 0$. Only in this situation can the ZECMs be effectively excited, and the injected photons will prefer locating on the corner site $(\mathbf{r}_\mathrm{c},\beta)$. Otherwise, we will have very low $R_{\mathbf{r}_\mathrm{c},\beta}$: If the ZECMs do not exist, we can only excite the spatially extended bulk or edge modes if $\Omega_{\mathrm{P}}$ is in the bulk or edge bands, leading to the dilution of the weight of SSPN on $(\mathbf{r}_\mathrm{c},\beta)$, or even can not effectively excite the lattice if $\Omega_{\mathrm{P}}$ is in the gap; if the ZECMs exist but we either do not pump the corner sites or pump the corner sites with $\Omega_{\mathrm{P}} \neq 0$, our pumping is not compatible with the ZECMs and thus can not effectively excite the ZECMs.
	%Empirically, if $R_{\mathbf{r}_\mathrm{c},\beta}$ exceeds $0.6$, it means that there are ZECMs, otherwise it does not exist shown in Fig.~\ref{Fig measurementfield}a.
	
	We perform our numerical simulation based on several recently reported experiments \cite{PhysRevApplied.15.014049,PhysRevApplied.15.064050,PhysRevLett.122.010501} and show our results in Fig.~\ref{Fig measurementfield}a-c. Here we choose $\kappa=0.03$ and set $\gamma=0.5$, $\lambda_i=1 $ following our previous analysis in Sec.~\ref{S} (see also Fig.~\ref{Fig butterfly}i). To detect the ZECMs, we set $\Omega_{\mathrm{P}}=0$ and choose $\mathcal{P}^{\dagger}\mathbf{a}=A_{1,1}+B_{1,8}+C_{8,1}+D_{8,8}$ (here we simulate the four-site pumping for better visualization. This choice does not affect our discussion and conclusion because the four single-site pumpings can be regarded as independent due to their distant spacing). We observe that $R_{\mathbf{r}_\mathrm{c},\beta}$ jumps dramatically at $\pi/2$ as shown in Fig.~\ref{Fig measurementfield}a. This jump is a clear evidence of the predicted topological phase transition induced by varying $\phi$ at $\phi=\pi/2$. In addition, our numerical calculation implies that $R_{\mathbf{r}_\mathrm{c},\beta}\textgreater0.7$ can serve as for the existence of the ZECMs. To better visualize the influence of the magnetic field $\phi$, we plot the SSPN of the whole lattice with topological non-trivial $\phi=2\pi/3$ and topological trivial $\phi=2\pi/10$ in Fig.~\ref{Fig measurementfield}b and c, respectively. The localization of the SSPN at $\phi=2\pi/3$ and its diffusion at $\phi=2\pi/10$ can be clearly identified.

	Here we offer a brief remark on the influence of the dissipation factor $\kappa$ on the localization of the photons. As the dissipation can also hinder the diffusion of the photons, we should discriminate whether the observed localization is caused by the appearance of ZECMs or by the dissipation of the lattice. Therefore, we consider the behavior of $R_{\mathbf{r}_\mathrm{c},\beta}$ and the SSPN under different $\kappa$. If their behavior is basically unchanged, we can conclude that the localization of the SSPNs is indeed caused by the appearance of ZECMs. We then perform the same numerical simulation with $\kappa=0.01$ and $\kappa=0.005$, and show the corresponding results in Figs.~\ref{Fig measurementfield}d-f and ~\ref{Fig measurementfield}g-i, respectively. With decreasing $\kappa$, the obtained results are largely unchanged, indicating that the observed localization is indeed caused by the appearance of ZECMs. Moreover, the jump of $R_{\mathbf{r}_\mathrm{c},\beta}$ at $\phi= \pi/2$ becomes increasingly sharp, and the SSPN at the topological trivial becomes more diffused. Therefore, smaller dissipation can help us to better identify the topological trivial/non-trivial regions. Meanwhile, as indicated in Fig.~\ref{Fig measurementfield}a, a moderate, currently achievable $\kappa=0.03$ is already sufficient for the discrimination the phase boundary.
	% So, we observe the SSPN of the corner is basically unchanged when the $\phi \in (\pi/2,3\pi/2)$ due to the presence of corner mode and There is significant change for the SSPNs when the $\phi \in (0,\pi/2)$ or $\phi \in (3\pi/2,2\pi)$ due to the absence of corner modes as shown in Fig.~\ref{Fig measurementfield}d and g. Meanwhile, compare with $\kappa=0.05$, the SSPN of the whole lattice with $\phi=2\pi/3$ and $\phi=2\pi/10$ is shown in ~\ref{Fig measurementfield}e,f,h and i. It is obvious that the SSPNs are basically unchanged when $\phi=2\pi/3$ in ~\ref{Fig measurementfield}e and h , and there is significant change for the SSPN for $\phi=2\pi/10$ as shown in Fig ~\ref{Fig measurementfield}f and i. Therefor, the localization of the SSPNs is caused by the corner model and the topological phase transition of propose cQED-2D SSH can be also clearly detected when $\kappa$ is relatively large in experiments.

	%\section{An aside: HOTI phase induced by anisotropy}

	\begin{figure*}[tbhp]
		\begin{center}
			\includegraphics[width=0.9\textwidth]{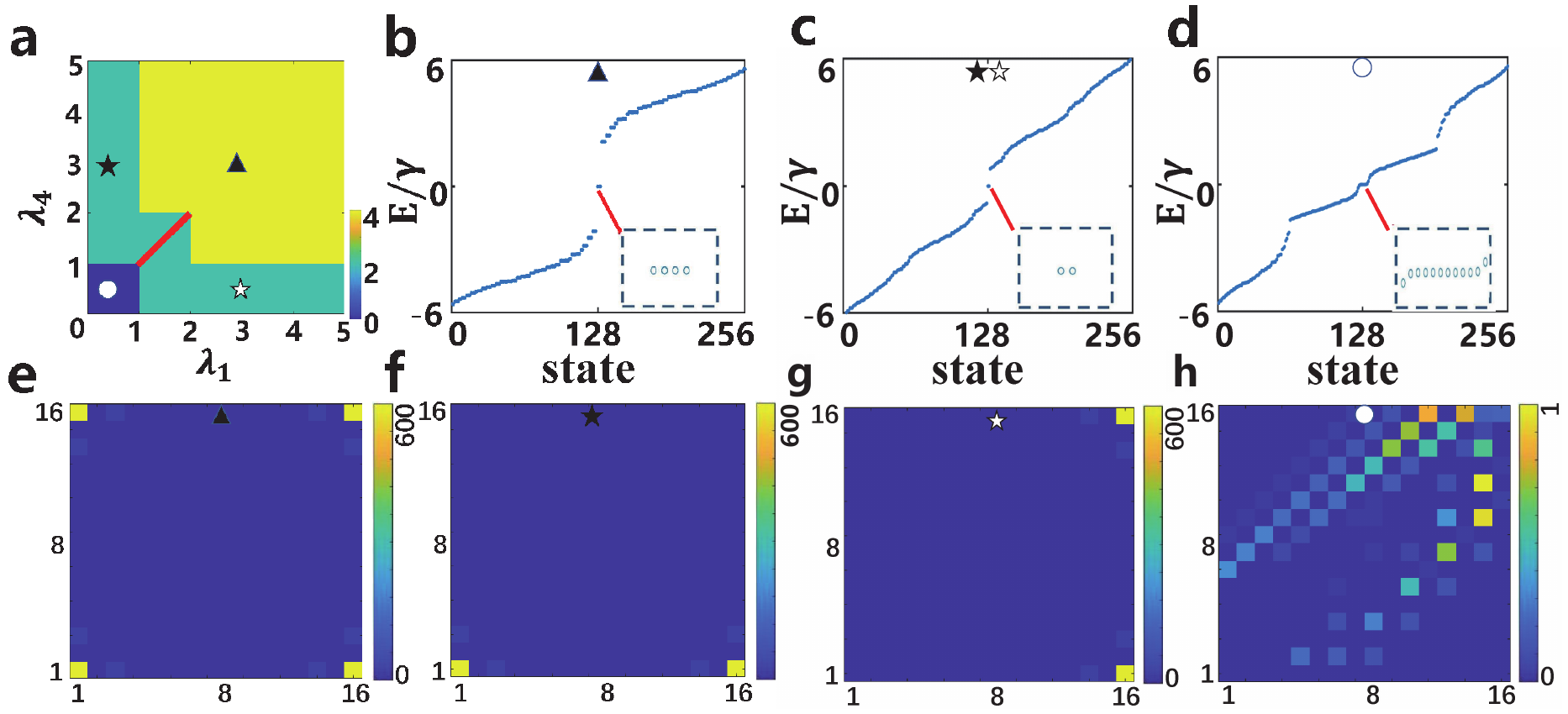}
		\end{center} 	
		\caption{\textbf{a} HOTI phase diagram of the 2D SSH lattice with $\phi=\pi$, $\gamma=1 $ and $\lambda_2=\lambda_3=3$. The phases are labelled by different colors indicating the number of the corner modes. In the blue, green, and yellow regions, the lattice have polarization vectors  $\boldsymbol{P} = (0,0), \boldsymbol{P}  = (0,1/2)$ (upper left) or $\boldsymbol{P}  = (1/2,0)$ (lower right), and $\boldsymbol{P}  = (1/2,1/2)$ and consequently zero, two, and four ZECMs, respectively. The spectra of the lattice at several representative points, labelled by triangle: $\lambda_1=\lambda_4=3$, star: $\lambda_1=3$, $\lambda_4=0.5$, hollowed star: $\lambda_1=0.5$, $\lambda_4=3$ and round: $\lambda_1=\lambda_4=0.5$, are shown in \textbf{b} - \textbf{d}, respectively. The spectrum behavior around the zero energy are detailed in the insets. The corresponding SSPN distributions under  $\kappa=0.03$, $\Omega_{\mathrm{P}}=0$, and $\mathcal{P}^{\dagger}\mathbf{a}=A_{1,1}+B_{1,8}+C_{8,1}+D_{8,8}$ are shown in  \textbf{e} - \textbf{h}, respectively.}
		\label{Fig corner}
	\end{figure*}
	
	Our proposed architecture can also be used to verify the anisotropy-induced HOTI predicted in Ref.~\cite{PhysRevB.98.205422}. Here we set $\phi=\pi,\gamma=1$, and $\lambda_2=\lambda_3=3$. The HOTI phase diagram of the lattice versus $\lambda_1$ and $\lambda_4$ is shown in Fig.~\ref{Fig corner}a. In particular, the yellow, green and blue regions label different polarization vectors \cite{Benalcazar61}  $\boldsymbol{P} = (0,0), \boldsymbol{P}  = (0,1/2)$ (upper left) or $\boldsymbol{P}  = (1/2,0)$ (lower right), and $\boldsymbol{P}  = (1/2,1/2)$, respectively. These different polarization vectors in turn correspond to different numbers and locations of the ZECMs. The energy spectrum of the lattice at several representative points are calculated and shown in Fig.~\ref{Fig corner}b-d: In the yellow region the lattice has four ZECMs locating at the four corners, while in the green region the lattice has two ZECMs, the location of which depending on $\lambda_1/\lambda_4$; However, the lattice has only extended bulk and edge modes but no isolated ZECMs in the blue region.% By taking into account the property of ZECMs, the system will have various types of topological phases. At present, because the hopping phase $e^{i \theta_{\mathbf{r}, \alpha}^{\mathbf{r}^{\prime}, \beta}}=\pm 1$ introduced by the magnetic field in Landau gauge, the topological phase can be characterized in terms of polarization $\boldsymbol{P}^{j}=$ $\left(P_{x}^{j}, P_{y}^{j}\right)$ as follows,

	We further numerically calculate the SSPN distribution of the lattice at these four representative points with $\kappa=0.03$, $\Omega_{\mathrm{P}}=0$, and $\mathcal{P}^{\dagger}\mathbf{a}=A_{1,1}+B_{1,8}+C_{8,1}+D_{8,8}$. In Fig.~\ref{Fig corner}e, we choose $\lambda_1=\lambda_4=3 $. We find that the SSPN is localized at four corners, indicating that there exist four ZECMs locating at the four corners. This is consistent with the spectrum shown in Fig.~\ref{Fig corner}b. Meanwhile, in Fig.~\ref{Fig corner}f with $\lambda_1=3$ and $\lambda_4=0.5 $, the resulting SSPN is localized only at two adjacent corner sites along the horizontal direction at the bottom. The SSPN distribution implies there exists only two ZECMs, which is consistent with Fig.~\ref{Fig corner}f. In Fig.~\ref{Fig corner}g, with inversely chosen $\lambda_1=0.5$ and $\lambda_4=3 $, the SSPN is localized at two adjacent corner sites along vertical direction. The comparison between Fig.~\ref{Fig corner}f and~\ref{Fig corner}g manifests their different polarization vectors. Finally, in Fig.~\ref{Fig corner}h we choose $\lambda_1= \lambda_4=0.5 $. Now the SSPN is no longer localized at the corner sites, implying that there is no ZECMs. These calculated SSPN distributions thus confirm the validity of our method, that is, we can extract the spatial and spectral information of the ZECMs by pumping the corner sites and detecting the corner sites and their few neighbors. Here we notice that our result is consistent with the previous theoretical work which states that the high order topological property is protected by the chiral symmetry and is not dependent on any spatial symmetry.  \cite{PhysRevB.98.205422}.

	%Experimentally, the scheme has already been used, where both the amplitude and the relative phase of a TLR coherent state were measured. It is worth nothing that we want to measure is only the expectation value $\left\langle \alpha_{\mathbf{r}}^{\dagger} \alpha_{\mathbf{r}}\right\rangle$ but not the detailed probability of the coherent steady-state projected to the Fock basis. It is weak requirement that greatly simplifies our measurement.

	\section{Conclusion and outlook}
	In conclusion, we have shown that it is not only possible, but also advantageous to implement and detect the HOTI phase transition of 2D SSH lattice induced by continuous magnetic field in SQC system.  Meanwhile, it is our feeling that we are still in the beginning of this research direction. Due to the flexibility of the cQED architecture, we can expect that other HOTI lattice configuration (e.g. kagome and honeycomb \cite{PhysRevLett.122.086804,PhysRevLett.120.026801}) can also be established by using our method. Moreover, the PFC method allows the further combination with many other mechanisms, including on-site Hubbard interaction \cite{2009,RevModPhys.85.299}, disorder, and non-Hermicity \cite{Ashida,doi:10.1080/00018732.2021.1876991}. In particular, the reduction of the fabrication error of cQED elements can be exploited to suppress and controllably introduce the mechanism of disorder \cite{UnderwoodDisorderPRA2012}. Another expectation comes from the fact that the effective photon-photon interaction, i.e. the nonlinearity of microwave photons can be incorporated into cQED system in a variety of manners, including the electromagnetically induced transparency \cite{HartmannCoupledCavity2006NP}, Jaynes-Cummings-Hubbard nonlinearity \cite{GreenTreeJCHNP2006}, and nonlinear Josephson coupling \cite{BourassaSQUIDCouplingPRA2012, LiebSQUIDCouplingNJP2012}. Our third perspective comes from the fact that the decoherence in superconducting quantum circuit can now be suppressed and efficiently controlled \cite{RN452}. This technique can be used to manipulate the gain and loss of the system, leading to the quantum simulation of non-Hermitian HOTI \cite{PhysRevLett.122.076801,PhysRevLett.123.073601}. Therefore, our further direction should be the interplay of the above mentioned mechanism in our proposed architecture, which will brings us into the realm of rich but less-explored physics.
	
	\acknowledgments
	We thank Y. H. Wu for helpful discussion. This work was supported in part by the National Science Foundation of China (Grants No.~11774114, No.~11874156, and No.~11874160).

%\bibliography{B}


\end{document}